# Magnetopause reconnection and indents induced by foreshock turbulence

Li-Jen Chen[1], Jonathan Ng[1,2], Yuri Omelchenko[3,4], Shan Wang[1,2]

[1] NASA Goddard Space Flight Center, Greenbelt, MD

[2] University of Maryland, College Park, MD

[3] Trinum Research, Inc., San Diego, CA

[4] Space Science Institute, Boulder, CO

**Key Points:**

- Foreshock turbulence can reach the magnetopause under a northward quasi-radial IMF with zero dawn-dusk component

- The turbulence can create large magnetic shear angles across the magnetopause, leading to local bursty reconnection

- Bombardments of the turbulence cause Earth-sized magnetopause indents under constant IMF and solar wind

Abstract

Based on global hybrid simulation results, we predict that foreshock turbulence can reach the magnetopause and lead to reconnection as well as Earth-sized indents. Both the interplanetary magnetic field (IMF) and solar wind are constant in our simulation, and hence all dynamics are generated by foreshock instabilities. The IMF in the simulation is mostly Sun-Earth aligned with a weak northward and zero dawn-dusk component, such that subsolar magnetopause reconnection is not expected without foreshock turbulence modifying the magnetosheath fields. We show a reconnection example to illustrate that the turbulence can create large magnetic shear angles across the magnetopause to induce local bursty reconnection. Magnetopause reconnection and indents developed from the impact of foreshock turbulence can potentially contribute to dayside loss of planetary plasmas.

1. Introduction

   Magnetopause reconnection plays a critical role in controlling energy input from the solar wind to the magnetosphere. The possibility that magnetic fluctuations generated in the foreshock may lead to magnetopause reconnection has been considered based on observations indicating that magnetosheath fields are not a simple compression of the interplanetary magnetic field (IMF) [Russell et al., 1997; Zhang et al., 1997]. Based on the fact that the statistical magnetopause locations for cone angles (between the IMF and the Sun-Earth line) greater and less than 45 degrees do not differ appreciably, the study by Zhang et al. [1997] concludes that the raised possibility is not confirmed. Recent measurements from the Magnetospheric Multiscale mission show Earth-radius-scale foreshock magnetic and density fluctuations approximately one order of magnitude larger than the respective upstream values under quasi-radial IMF (cone angles < 30 degrees) conditions [Chen et al., 2021]. These intense fluctuations often contain strong $B_z$ (in Geocentric Solar magnetic (GSM) coordinates) and $B_y$ components. If these fluctuations can reach the magnetopause, they can create conditions conducive to reconnection. In this paper, based on global hybrid simulation results, we predict that foreshock fluctuations in the form of strong $B_z$ and density fluctuations can reach the magnetopause and lead to reconnection as well as Earth-sized indents.

   To single out the effects of foreshock fluctuations, we design our simulation to have a quasi-radial IMF with a weak northward ($B_z>0$) and zero dawn-dusk ($B_y=0$) components. When the IMF is quasi-radial, foreshock waves excited due to reflected ions interacting with the incoming solar wind are the strongest, permeate the largest volume of the subsolar region, and can most strongly influence the inner magnetosphere [e.g.,

Russell et al.,1983; Takahashi et al., 2021, and references therein]. In the aspect of reconnection, extensive statistical studies show that magnetopause reconnection occurs for shear angles (angle between the magnetosheath and magnetosphere magnetic fields across the magnetopause) as low as ~30 degrees but not lower ~~that~~ than that [e.g., Trattner et al., 2017a, 2017b; Fuselier et al., 2017]. Our choice of the IMF $B_z>0$ and $B_y=0$ renders magnetic reconnection improbable at the subsolar magnetopause, where the shear angle is ~ 0, without foreshock fluctuations altering the magnetosheath fields.

2. The global hybrid simulation

To address the posed open question, we employ the state-of-the-art hybrid code HYPERS [Omelchenko and Karimabadi, 2012; Omelchenko et al., 2021, and references therein] that treats ions as particles and electrons as a massless fluid. The hybrid approach is essential to describe the interaction of the solar wind with particles reflected from the bow shock [e.g., Burgess, 2005; Karimabadi et al., 2014] and capture the ensuing magnetopause impact. The dynamics generated by this interaction will be referred to as foreshock turbulence hereafter.

We use a stretched mesh in the simulation to meet in part the multi-scale challenge of the problem. The mesh resolution is one solar wind ion inertial length ($d_i$) in the central domain containing the magnetosheath, magnetopause, cusps, and a part of the foreshock. Outside the central domain, the cells are stretched exponentially in all dimensions, reaching the maximum length at the outer boundary (~12 $d_i$). The system size is $300(x) \times 500(y) \times 500(z)$ cells, covering a spatial domain of ~ $700(x) \times 1400(y) \times 1400(z)\ d_i$. The central domain occupies the region $x$=[0, 175], y=[-

125, 125], and z=[-125, 125] (in units of $d_i$). Our magnetopause standoff distance is 120 $d_i$, which can be scaled to the typical magnetopause distance 10 $R_E$. The coordinates are given in GSM. The density and magnetic field are normalized by their upstream values; velocity components are normalized by the upstream Alfvén speed.

We integrate particle trajectories and discrete fields using asynchronous adaptive time steps automatically determined by the simulation based on local physics and mesh properties [Omelchenko et al., 2021, and references therein]. This time adaptive approach addresses the following critical computational demands: (1) to resolve ion motion over a range of background magnetic field strengths spanning orders of magnitude, and (2) to resolve dynamic variations involved in turbulence and reconnection where the local time scales are not known *a priori*.

To capture magnetic reconnection in a hybrid simulation which lacks electron kinetic physics, plasma resistivity needs to be controlled and modeled adequately. Our hybrid simulation imposes the Chodura resistivity (e.g., Sgro and Nielson [1976]; Milroy and Brackbill [1982]), a semi-empirical model with dependence on the current density and number density. The Chodura resistivity has been employed in simulations of reconnection in field-reversed configurations, showing agreement with experiments (e.g., Omelchenko [2015]; Kayama, [2012]). We have benchmarked our resistivity model against full particle-in-cell simulations of established configurations [e.g., Chen et al., 2012] and reproduced electron-scale reconnecting current sheets.

The simulation conditions are as follows: the Alfvén Mach number $M_A=V_{sw}/V_A=8$, the electron and ion beta $\beta_e=\beta_i=0.5$, the IMF $B_z > 0$, $B_y = 0$, and cone angle 10 degrees, and $c/V_A = 8000$. The Earth's dipole tilts 11.5 degrees towards the Sun such that the Northern hemisphere is in the summer. The simulation is run for $437\omega_{ci}^{-1}$, where $\omega_{ci}$ is the ion cyclotron frequency based on the IMF.

The boundary conditions are (1) inner boundary: a particle-absorbing, perfectly conducting sphere with a radius $r = 50 d_i \sim 4 R_E$ surrounded by a dense cold plasma to model the ionospheric shorting out of electric field, (2) outer $x$ boundary: a steady solar wind inflow along -$x$ with density, temperature, and velocity identical to those of the initial plasma at the positive $x$ boundary; outgoing solar wind particles escape through the outflow (negative $x$) boundary and backstreaming particles are allowed to escape through the inflow boundary, (3) outer y and $z$ boundaries: semi-reflecting conditions which reflect drifting Maxwellian (solar wind) particles and absorb other particles.

3. Simulation results

First we demonstrate that foreshock turbulence leads to Earth-size indents at the magnetopause. The IMF and solar wind are both steady with no variations. Yet the 3D magnetopause configuration changes significantly (Figure 1): indents (marked by the white ovals) occur at different parts of the magnetopause frequently. The two shown examples at time $t_1$ & $t_2$ are separated by $12.5\omega_{ci}^{-1}$. The surface represents the magnetopause (defined by the local density $N = 1.5 N_{sw}$ within a radius of 145 $d_i$ from the point [0, 0, -15] in GSM coordinates, where $N_{sw}$ is the density of the solar wind; this region constraint ensures that only the magnetosheath and the magnetosphere are searched to find the magnetopause. This operational definition is approximately consistent with the definition based on the open-closed field line boundary – see Figure 3) colored with the magnetic field magnitude |B|. The large indent shown at $t_2$ lasts for a total of ~ $19\omega_{ci}^{-1}$ (~1 minute using 4 nT IMF). In this simulation, an Earth-sized magnetopause indent can last for $4\text{-}20\omega_{ci}^{-1}$, and one to multiple indents often occur after the magnetosphere is well established. Furthermore, an indent does not necessarily occur after a bulge, in contrast to the scenario indicated for the magnetopause compression due to the rim of a foreshock cavity [Lin and Wang, 2005]. We emphasize that the indent is a result of the dynamical magnetopause evolution in response to the Earth-sized fluctuations from the foreshock on the time scale of minutes, and is not based on a comparison with the magnetopause locations under other IMF orientations [Zhang et al., 1997] or those predicted by MHD [Samsonov et al., 2012] and empirical models [Merka et al., 2003; Dusik et al., 2010; Suvorova et al., 2010].

The magnetic field magnitude |B| and field lines illustrate foreshock waves, amplified magnetic structures, and the magnetopause indent (Figure 2; same colorbar as that in Figure 1). The magenta circle marks the location of a fixed sphere from which field lines are traced until they reach one of the boundaries. At $t_1$, the magenta region (slightly inside of the magnetopause) holds only closed field lines. At $t_2$, the same region carries field lines connected to the inner boundary at the northern cusp and open to the dayside solar wind (two examples are shown), implying that the magnetopause has moved earthward into the magenta circle. This is consistent with the fact that the boundary between the red and gray contours of |B| has moved into the magenta circle at $t_2$. The dimension of the magnetopause indent is ~ 0.7 $R_E$ (~8 $d_i$ based on Figures 2 & 4) along $x$ and a few $R_E$ along $z$.

A global picture of how foreshock turbulence increases the likelihood of magnetopause reconnection equatorward of the cusps is presented through analyzing shear angles of the magnetosheath field (modified by the turbulence from the foreshock) and magnetosphere field across the magnetopause. As our key interest is on the magnetic fields for latitudes lower than the cusps, here we define the magnetopause $x$ location at each (y, z) point by the first open field line, and name it $x_{mp,open}$ (Figure 3a). Operationally, we trace field lines starting from the closed-field-line region, and increase $x$ until we find the first field line that is at least open on one end. The magnetopause location at $t_2$ defined this way exhibits a similar large-scale indent as that shown in Figure 1. The shear angle is computed as $acos\left(\frac{\boldsymbol{B}_{sh} \cdot \boldsymbol{B}_{sp}}{|\boldsymbol{B}_{sh}||\boldsymbol{B}_{sp}|}\right)$ (Figure 3b), where $\boldsymbol{B}_{sh}$ and $\boldsymbol{B}_{sp}$ are the magnetic fields at 3 $d_i$ from $x_{mp,open}$ on the magnetosheath and magnetosphere sides, respectively. Large shear angles (>~ 90 degrees) populate the indent region with z < 0 and y ~ [-20, 20].

One example of magnetopause reconnection in the indent region is shown to be associated with an ion flow reversal (Figures 3c-d). Four field lines representing the magnetosheath inflow (green), magnetosphere inflow (yellow), north exhaust (blue), and south exhaust (magenta) meet in the proximity of an ion flow $V_z$ reversal (interpreted as the reconnection outflows) and a negative $B_z$ region extending about 20 $d_i$ along the magnetopause. The non-ideal electric field is enhanced at the location with correlated magnetic field and flow reversals. This example presents simulation evidence for magnetopause reconnection induced by amplified foreshock waves that generate the negative $B_z$ and finite $B_y$. The reconnection lasts for several $\omega_{ci}^{-1}$, approximately the duration when the negative $B_z$ region is right against the magnetopause. This type of bursty reconnection is a product of the impact of foreshock turbulence on the magnetopause.

The amplification and evolution of foreshock waves, as they are convected from the foreshock through the magnetosheath toward the magnetopause, are presented in sample 2D views (Figure 4). Ions reflected from the shock interact with the incoming solar wind, and generate the ultra low frequency (ULF) electromagnetic waves. These waves are amplified to supply negative $B_z$ (Figure 4, top panels) to the magnetopause. The waves often grow into isolated structures with enhanced |B|, density N, and $B_z$ as well as $B_y$ of both signs, consistent with the solitary magnetic structures observed by the MMS spacecraft and reproduced by full PIC simulations [Chen et al., 2021] (also known as isolated Short-Large-Amplitude-Magnetic-Structures, or SLAMS [Schwartz et al., 1992]). For example, the solitary density structures marked by arrows in the density profiles (Figure 4, bottom panels) have corresponding |B| enhancements larger than twice the background magnetic field amplitude, satisfying the SLAMS criteria listed in Schwartz et al. [1992]. Regions of strong southward $B_z$ resulting from amplified foreshock turbulence are convected through the magnetosheath to enable subsolar magnetopause reconnection otherwise not expected to occur under a steady IMF with positive $B_z$ and zero $B_y$.

The magnetopause indent at $t_2$ (Figure 1, right panel) resulted from a series of bombardments of enhanced density and magnetic field structures. The density profiles indicate that the density further builds up at $t_2$ near the subsolar magnetopause (Figure 4, bottom panels), and the magnetopause is pushed further inward at $t_2$ compared to that at $t_1$. Neither preceding density cavities (Figure 4, bottom panels) nor localized flow enhancements are observed. We therefore conclude that an indent like this - commonly observed in our simulation - is not associated with foreshock/magnetosheath cavities (density depression preceding a density rim – see [Lin and Wang, 2005; Omidi et al., 2016]), nor magnetosheath high-speed jets (HSJs defined in [Plaschke et al., 2013] as localized earthward flow enhancements with at least half of the solar wind dynamic pressure).

4. Summary and discussion

In summary, based on global hybrid simulation results discussed in this paper, we predict that foreshock turbulence can penetrate through the magnetosheath, and lead to magnetopause reconnection as well as indents under a constant solar wind and northward quasi-radial IMF with zero dawn-dusk component. The turbulence supplies Earth-size regions of highly enhanced densities and magnetic fields $B_{yz}$, corresponding to solitary magnetic structures [Chen et al., 2021] or SLAMS [Schwartz et al., 1992], to the magnetopause, inducing reconnection and indents equatorward of the cusps. Turbulence-induced magnetopause reconnection are local and bursty in nature, as the spatial and temporal scales of the magnetosheath fields being reconnected are influenced by the parent foreshock processes.

Magnetopause reconnection induced by foreshock turbulence presents a new scenario of dayside reconnection when the IMF clock angle (between the IMF $B_{yz}$ and $z_{GSM}$) is ~0, as in the simulated case. The other known scenarios under northward IMF conditions are: (1) reconnection of IMF with open cusp field lines (lobe reconnection), and (2) reconnection of IMF with closed field lines that thread through the nightside of the magnetosphere (non-lobe reconnection), as illustrated in Fuselier et al. [2018]. We show that foreshock turbulence makes possible for local bursty reconnection to occur near the subsolar magnetopause by creating large shear angles between the magnetosheath and the magnetosphere fields.

Our simulation results indicate that the impact of foreshock turbulence on the magnetopause can take forms other than high-speed magnetosheath jets [Plaschke et al., 2013; Hietala et al., 2018, and references therein], foreshock/magnetosheath cavities [Lin and Wang, 2005; Omidi et al., 2016], or spontaneous hot flow anomalies [Omidi et al., 2016]. The Earth-sized indents (Figure 1) are primarily due to bombardments of the magnetopause by structures with enhanced densities and magnetic fields. In our simulation, magnetopause bulges similar to those reported earlier [Lin and Wang, 2005; Omidi et al., 2016] occur as well, although not necessarily preceding the indents. Effects of indents and bulges may be averaged out, and this may be why the nose of the magnetopause location for cone angles less than 45 degrees does not statistically exhibit a significant difference from that for cone angles greater than 45 degrees [Zhang et al., 1997]. This consideration indicates the importance of case studies to validate our prediction of foreshock turbulence penetrating through the magnetosheath and inducing magnetopause reconnection and indents.

Turbulence-induced magnetopause indents and reconnection can potentially lead to dayside loss of planetary plasmas. Our simulation results indicate that Earth-sized indents are common and often occur at multiple magnetopause locations. When magnetopause indents intercept with the drift shells of energetic particles, referred as magnetopause shadowing in the literature [e.g., Sorathia et al., 2017], particles can escape from the dayside magnetosphere even in the absence of magnetopause reconnection. These particle losses may appear bursty or continuous since multiple channels for ion escape may overlap in time and space. Bombardments of foreshock turbulence on the magnetopause at various locations may explain bursts of $O^+$ ions (~1 minute duration for each burst, comparable to the indent lifetime) observed upstream close to the bow shock [Möbius et al., 1986]. When the bombardments at different locations overlap in time, they may account for longer duration $O^+$ events as far as L1 [Posner et al., 2002, 2003] under quasi-radial IMF. With the added power of reconnection, closed field lines can be open to the dayside solar wind (as shown in Figure 2), leading to direct leakage of ionospheric and ring current ions, potentially reaching L1 given the nearly radial IMF. Further observation and simulation studies are needed to assess the effectiveness of this dayside planetary plasma loss mechanism.

Quasi-radial IMF conditions are not rare. A statistical survey covering more than a solar cycle indicates that the occurrence rate for long-duration quasi-radial IMF (cone angles less than 25 degrees and lasting for >= 4 hours) is 10-15%, and this rate is independent of the solar cycle [Pi et al., 2014]. For the foreshock kinetic instabilities to occur and generate Earth-sized turbulent structures that can impact the magnetosheath and magnetopause, the quasi-radial IMF only needs to last for a few minutes based on MMS observations [Chen et al., 2021], and hence the occurrence rate for quasi-radial IMF with potential magnetopause impact is expected to be higher than the above value. Our deliberate choice of a constant quasi-radial IMF with a weak $\boldsymbol{B}_z > 0$ and zero $\boldsymbol{B}_y$ allows the impact of foreshock turbulence on magnetopause dynamics to be unambiguously identified. A quasi-radial IMF does not need to take this special $B_{yz}$ orientation to have the same effectiveness on the magnetopause.

Recognizing the limitations of the simulation is important for envisioning future steps to advance the subject research. The typical terrestrial magnetopause standoff distance is approximately 600 $d_i$, five times of that used in our simulation. While our simulation results are applicable to a magnetosphere smaller than the Earth's, the effects of foreshock turbulence may be exaggerated for terrestrial applications. For example, (1) foreshock turbulence may be dissipated before reaching the Earth's magnetopause, (2) the time duration for enhanced magnetic shear (and hence current density) may be too brief for reconnection to develop, and (3) even though we have benchmarked our resistivity model by reproducing laboratory experimental results and electron-scale reconnection current sheets formed in full PIC simulations, how well this resistivity model describes magnetopause reconnection awaits observations to judge. Multi-spacecraft measurements from the magnetopause and upstream solar wind with northward quasi-radial IMF (negligible $B_y$) will provide a stringent validation of our prediction.

Acknowledgments

The research is supported in part by the MMS mission, DOE grants DESC0016278, DESC0020058, NSF AGS-1619584, AGS-2010231, and NASA 80NSSC18K1369 and 80NSSC19K0838. The authors thank Johnny Zhang, Michael Heinsohn, Nancy Carney, and other NASA Advanced-Supercomputing and High-End-Computing team members for their professional support to make the simulation possible. Simulation data are available at http://doi.org/10.5281/zenodo.4394527.

**Captions**

**Figure 1.** Variations of the 3D magnetopause configuration under a steady quasi-radial IMF with $B_z>0$ and $B_y=0$: Earth-sized indents (marked by the white ovals) at different parts of the magnetopause at two different times $t_1$ & $t_2$ ($t_2 = t_1 + 12.5\omega_{ci}^{-1}$; $t_1$ is $266\omega_{ci}^{-1}$ after the start of the simulation). The surface represents the magnetopause (defined by the local density $N = 1.5\ N_{sw}$, where $N_{sw}$ is the density of the solar wind) colored with the magnetic field magnitude |B|. The magnetopause indents are caused by bombardments of penetrating foreshock turbulence, as both the solar wind and the IMF are constant.

**Figure 2**. Magnetic field amplitude and field lines showing foreshock waves, amplified magnetic structures, and the indent. The magenta circle marks the location of the sphere (radius 5 $d_i$, centered at [117.3, -0.5, -8.7]) from which field lines are traced until they reach one of the boundaries. At $t_2$, field lines connected to the inner boundary at the northern cusp have been open to the dayside solar wind (left boundary). The color presents |B| (same colorbar as that in Figure 1). The 2D x-z domain is taken at $y = 0$. The white sphere represents the inner boundary at ~4 $R_E$ from Earth's center, and the blue circle the Earth in approximate proportion. The cyan box marks the domain to be presented in Figure 4. Enhanced magnetic field fluctuations are seen reaching the magnetopause both equatorward and tailward of the cusp. Note that the IMF and solar wind are kep constant, and hence all the fluctuations are due to foreshock turbulence.

**Figure 3.** Demonstration of foreshock turbulence creating large shear angles across the magnetopause and an example of turbulence-induced reconnection near the subsolar magnetopause. (a) The $x$ locations of the magnetopause ($x_{mp,open}$) defined by the first open field line. (b) Magnetic field shear angles calculated as the angle between the magnetic fields at 3 $d_i$ from $x_{mp,open}$ on the magnetospheric and magnetosheath sides. (c) Four field lines representing the magnetosheath inflow (green), magnetosphere inflow (yellow), north exhaust (blue), and south exhaust (black) over-plotted on $B_z$. (d) The four field lines meet in the proximity of an ion flow $V_z$ reversal, interpreted as the reconnection outflow. The 2D $x$-$z$ domain is taken at $y = 6$. All data are from $t_2$.

**Figure 4.** Foreshock waves are amplified to form nonlinear structures with strong southward $B_z$. Regions with negative $B_z$ are convected to the magnetopause to create large shear angles across the magnetopause. Bottom panels show the corresponding density profiles on the same $x$-$y$ plane. The frames are consecutive and adjacent frames are separated by $3.125\omega_{ci}^{-1}$. The black arrow represents the direction of increasing time. Arrows in the density profiles (bottom panels) indicate examples of solitary magnetic structures [Chen et al., 2021]. The evolution of a specific structure is tracked with arrows of the same color. The 2D $x$-$y$ domain is taken at $z = 0$.

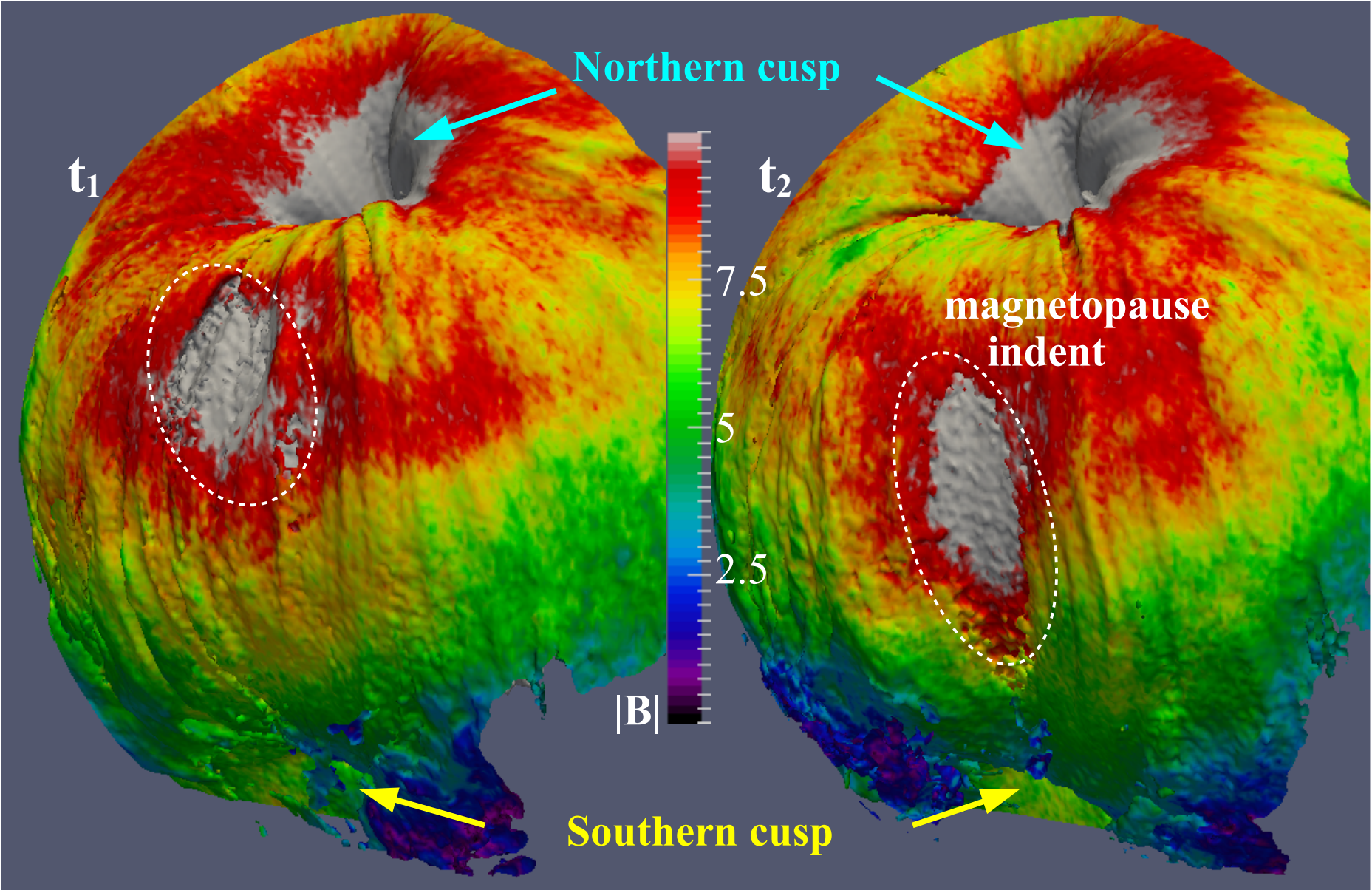

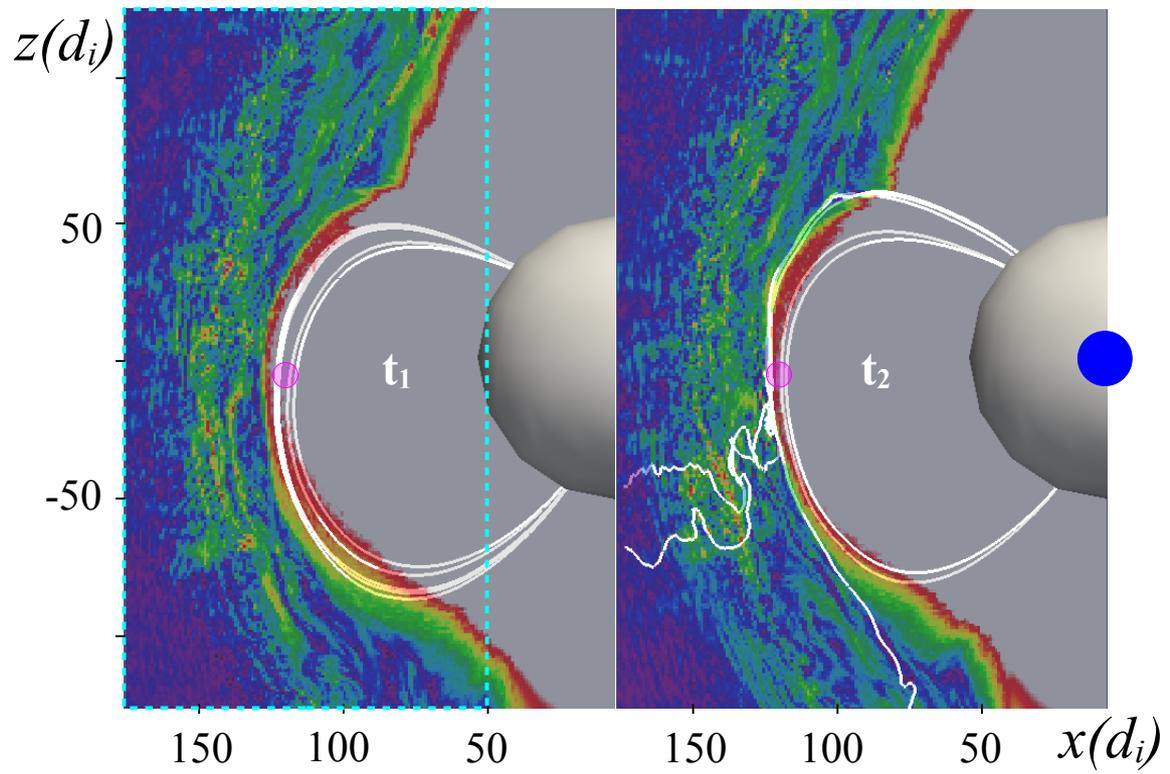

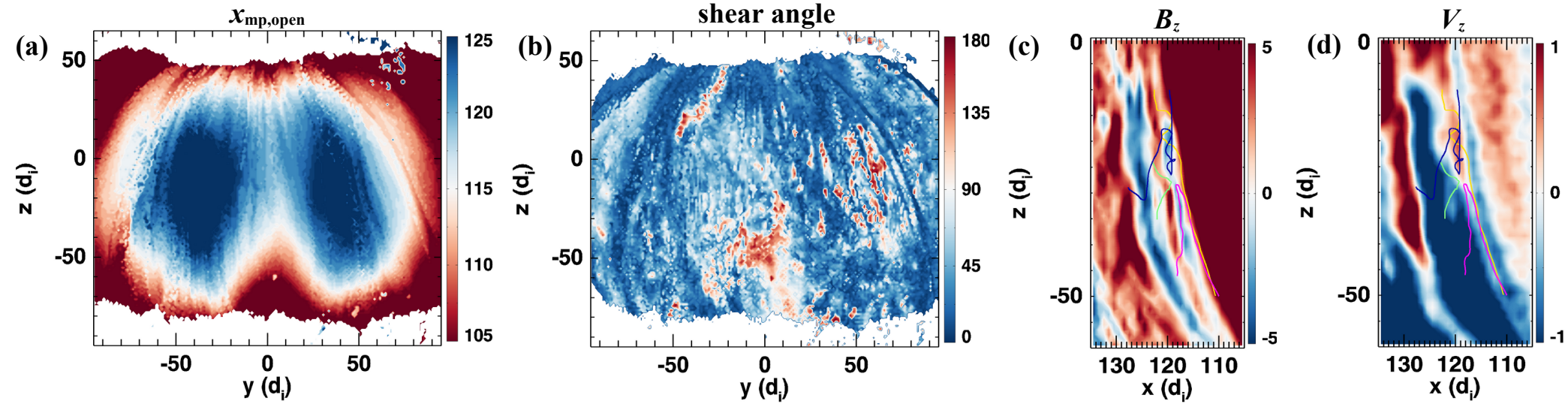

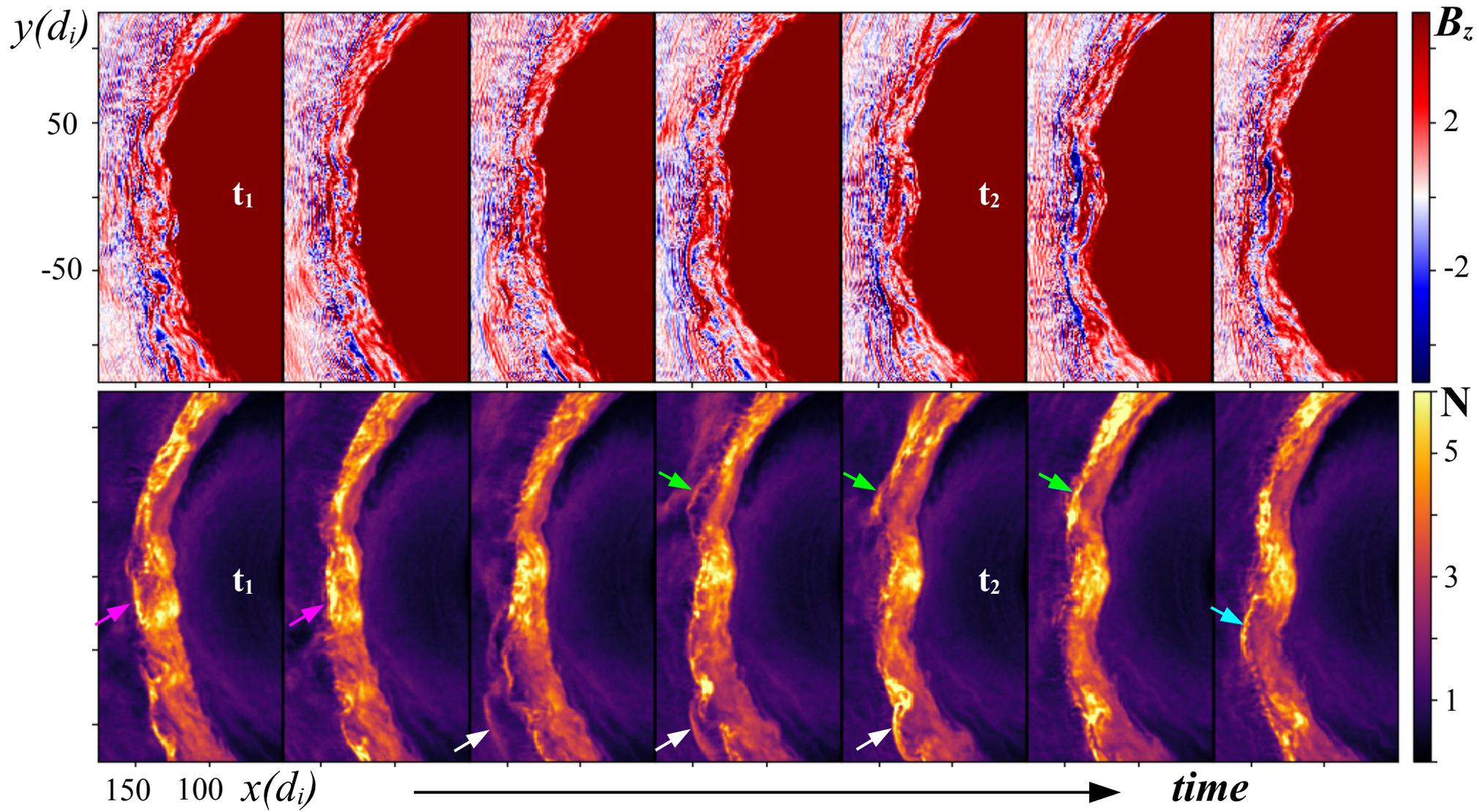